\documentclass[aps,pre,amsmath,amssymb,reprint,superscriptaddress,twocolumn,showpacs]{revtex4-1}

 \usepackage{graphicx}
 \usepackage{amsmath}
 \usepackage{amsfonts}
 \usepackage{amssymb}
 \usepackage{pstricks}
 \usepackage{psfrag}
 \usepackage{epsfig}
 \usepackage{changes}
 \usepackage[ansinew]{inputenc}
\usepackage{mathptmx}
\usepackage{helvet}
\usepackage{textcomp}
\usepackage{ulem}

\begin{document}


\title{{\color{black} Inertial clustering} and emergent phase separation of spherical spinners}



\author{Zaiyi Shen}
\affiliation{Univ. Bordeaux, CNRS, LOMA (UMR 5798), F-33405 Talence, France}
\author{Juho S. Lintuvuori}
\email[]{juho.lintuvuori@u-bordeaux.fr}
\affiliation{Univ. Bordeaux, CNRS, LOMA (UMR 5798), F-33405 Talence, France}



\begin{abstract}
We study the hydrodynamics of {\color{black} spherical spinners suspended in a Newtonian fluid} at inertial regime.
{\color{black} We observe a spontaneous condensation of the spinners into particle rich regions}, at low but finite particle Reynolds numbers and volume fractions.
The particle clusters have a coherent internal dynamics. The spinners form colloidal vortices surrounded by the fluid depleted of the particles. The formation of vortices is observed both in periodic simulation box and when the spinners are confined between two flat walls.
The stabilisation of the observed states relies only on hydrodynamic interactions between the spinners and requires a finite amount of inertia. The observations pave the way for the realisation of 3-dimensional spinner materials, where coherent structures and collective dynamics arise only from the rotational motion of the constituents.  \end{abstract}


\maketitle


\section{Introduction}
Creating dynamic structures from motile units is ubiquitous in {\color{black} a} natural world.
A topical example, in a micrometer length scale, is provided by active materials~\cite{marchetti2013hydrodynamics}{\color{black}. The} motile units, such as bacteria~\cite{cates2012diffusive} or phoretic Janus colloids~\cite{moran2017phoretic}, form ordered states.
Another example is provided by torque-driven colloidal particles~\cite{snezhko2016complex}, where the particle rotation is achieved by either magnetic{\color{black}~
\cite{grzybowski2000dynamic,grzybowski2002dynamic,martinez2018emergent,tierno2008controlled,driscoll2017unstable,kokot2018manipulation,kaiser2017flocking,yan2015colloidal,yan2012linking,sing2010controlled,steimel2014artificial}}, electric~\cite{bricard2013emergence,bricard2015emergent}, acoustic~\cite{zhou2017twists,sabrina2018shape}  or optical fields~\cite{paterson2001controlled,friese1998optical,padgett2011tweezers,yang2018systems}.
In a typical experimental realisation, the particle rotation is converted to a translational rolling motion due to a presence of a surface~\cite{martinez2018emergent,tierno2008controlled,driscoll2017unstable,kokot2018manipulation,kaiser2017flocking,bricard2013emergence,bricard2015emergent,gotze2011dynamic}.
The formation of various dynamical states, such as flocking~\cite{bricard2013emergence,kaiser2017flocking},  complex motile structures~\cite{driscoll2017unstable,tierno2008controlled}  and vortical motion both with~\cite{bricard2015emergent,kaiser2017flocking} and without confinement~\cite{kokot2018manipulation} has been observed by field-actuated particles.

Another possibility is provided by systems, where the dynamics of the individual building blocks is purely rotational.
Nature's examples of this include the dancing of Volvox~\cite{drescher2009dancing}  and the formation of vortex arrays~\cite{petroff2015fast} of spinning bacterial cells.
Previous studies of artificial colloidal spinner materials~\cite{fily2012cooperative,yeo2015collective,spellings2015shape,nguyen2014emergent,van2016spatiotemporal,shen2019hydrodynamic,goto2015purely} have concentrated either to very low Reynolds numbers or to 2 dimensions (2D).
{\color{black} At vanishing Reynolds numbers and at high area fractions simulations have predicted a phase separation of binary mixtures~\cite{yeo2015collective,spellings2015shape,nguyen2014emergent}}, the emergence of edge currents~\cite{van2016spatiotemporal} as well as the stabilisation of 2D crystals~\cite{shen2019hydrodynamic}, while the presence of an odd viscosity has been predicted for chiral active fluids~\cite{banerjee2017odd}.
Experiments of circular disks spinning  on a gas-liquid interface at finite Reynolds numbers, demonstrated a dynamic ordering arising from the competition between magnetic attraction and hydrodynamic repulsion~\cite{grzybowski2000dynamic,grzybowski2002dynamic}. While attractive interactions has been observed for inertial spinners in a dense passive media{\color{black}~\cite{aragones2016elasticity,steimel2016emergent}} and {\color{black} ordered structures have been predicted for spinning disks at finite Reynolds numbers in 2D~\cite{goto2015purely}. {\color{black} The effects of inertia in 3-dimensional (3D) spinner solutions are currently unknown.}

We consider a simple 3D  spinner system, {\color{black}consisting} of spherical particles suspended in a Newtonian fluid and subjected to torques.} {\color{black} Our results demonstrate that the initially uniform particle density is unstable: a spontaneous condensation of particle rich and particle poor domains from the initially uniform distribution of the spinners is observed.} {\color{black} When both the particle volume fraction and Reynolds number are low but finite, the spinners spontaneously organise into colloidal vortices surrounded by a pure fluid. We demonstrate that} the observed emergent phase separation and the collective motion originates solely from the hydrodynamic interactions between the spinners and requires a finite amount of inertia.

\section{Methods}
We use a lattice Boltzmann method (LBM)~\cite{succi2001lattice}, to study {\color{black} the dynamics of} suspensions of rotationally driven spherical particles. {\color{black}The LBM was used to solve the quasi-incompressible Navier-Stokes equation for the fluid flow~{\color{black}\cite{succi2001lattice}}.
The no-slip boundary condition on the particle surface is realised by bounce back on links methods~\cite{ladd1,ladd2} which can be modified to take into account the movement of the particles~\cite{nguyen2002lubrication}. We considered a density matched solution $\rho=\rho_{\mathrm{fluid}}=\rho_{\mathrm{particle} }$. We set the LBM lattice spacing $\Delta x = 1$, time unit $\Delta t = 1$ and density $\rho = 1$, as customary in LBM.
The spinners were modelled as spherical particles, with radius $R = 6\Delta x$ {\color{black}($R = 2.1\Delta x$ in Fig.~4)}, with a very short range repulsion between them to avoid particle-particle overlaps~\cite{shen2018hydrodynamic,shen2019hydrodynamic}.}
A constant torque $T$ (Fig.~\ref{tornado}a) is applied on each particle. This leads to a spinning motion around a unique axis ($X$) which gives the (particle) vorticity direction (Fig.~\ref{tornado}a). {\color{black} For a small particle Reynolds number (Re) an isolated spinner reaches a steady state rotational frequency $\omega_0 = T/8\pi\mu R^3$}, {\color{black} where $\mu$ is the dynamic viscosity of the fluid}.
{\color{black} When Re is increased a deviation from $\omega_0$ is expected  due to inertial effects with the observed $\omega < \omega_0$~\cite{bickley1938lxv,liu2010wall,collins1955steady}.}
For the $\mathrm{Re} \lesssim 20$ considered in this work the deviation is reasonably small (Fig.~\ref{validation}), thus $\omega_0$ was used to calculate the particle Reynolds numbers used in the text. The particles are placed in a three dimensional periodic rectangular box with a square cross section in $YZ$ plane and squashed along the vorticity direction $X$ (Fig.~\ref{tornado}b) ({\color{black} Two parallel solid walls are added in $YZ$ plane in Fig.~\ref{wall}a, b and c}).

The dynamical state of the system is characterised by the particle Reynolds number $\mathrm{Re}=\tfrac{\rho\omega_0 R^2}{\mu}$ and the particle volume fraction $\phi = N\tfrac{4/3\pi R^3}{L_XL_YL_Z}\times 100$\%, where  $L_{X|Y|Z}$ are the simulation box lengths.

Assuming a particle with a radius $100\mu$m and using the kinematic viscosity of water $10^{-6}$m$^2$/s, a $\mathrm{Re}=1$ would require a spinning frequency $\omega_0=100$Hz. Using these values we can map a single simulation time unit to $\Delta t\approx 5\times 10^{-5}$s. A typical simulation run consisting of $10^{6}$ LB steps, corresponds to $50$s in real time.

\section{Results}

\begin{figure}
\centering
\includegraphics[width=1\columnwidth]{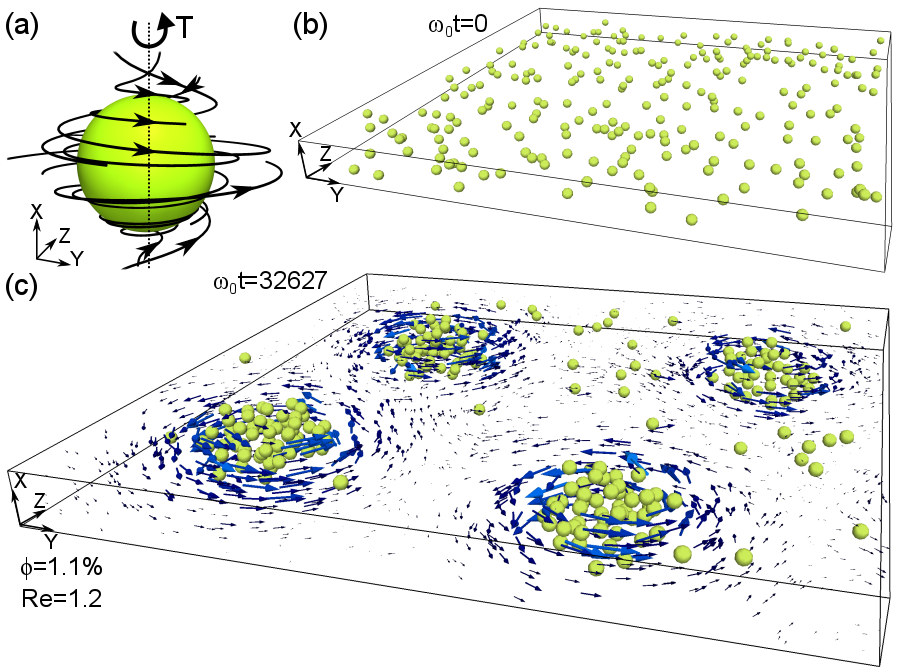}
\caption{\footnotesize {Rotating clusters formed by spinning spherical particles.} (a) The streamlines around a rotating particle driven by a constant torque around the vorticity axis $X$ at $\mathrm{Re}\approx10$. (b) The simulations are started with a random particle positions  in a rectangular box with a square cross section in $YZ$ plane.  (c) Observed formation of rotating clusters at a volume fraction $\phi \approx 1.1$\% with $\mathrm{Re}\approx1.2$. (The yellow spheres mark the particles and the blue arrows show the fluid flow field $\mathbf{u}$. A periodic simulation box  of $8R\times 108R\times 108R$ was used.) \label{tornado}}
\end{figure}

When starting from a random initial positions (Fig.~\ref{tornado}b) we observe that  the hydrodynamic coupling between the spinners leads to {\color{black} the formation of small clusters  and eventually to a phase separation ({\color{black} Fig.~\ref{tornado}c}). The particles spontaneously organise themselves into rotating clusters surrounded by clear fluid when {\color{black} $\phi\sim 1 \%$ and $\mathrm{Re}\sim 1$} (see e.g. Fig.~\ref{tornado}c for $\phi\approx 1.1$\% and $\mathrm{Re}\approx 1.2$; {\color{black} as well as Movie 1 in ~\cite{SM}}).

\begin{figure}[tbhp!]
\centering
\includegraphics[width=1\columnwidth]{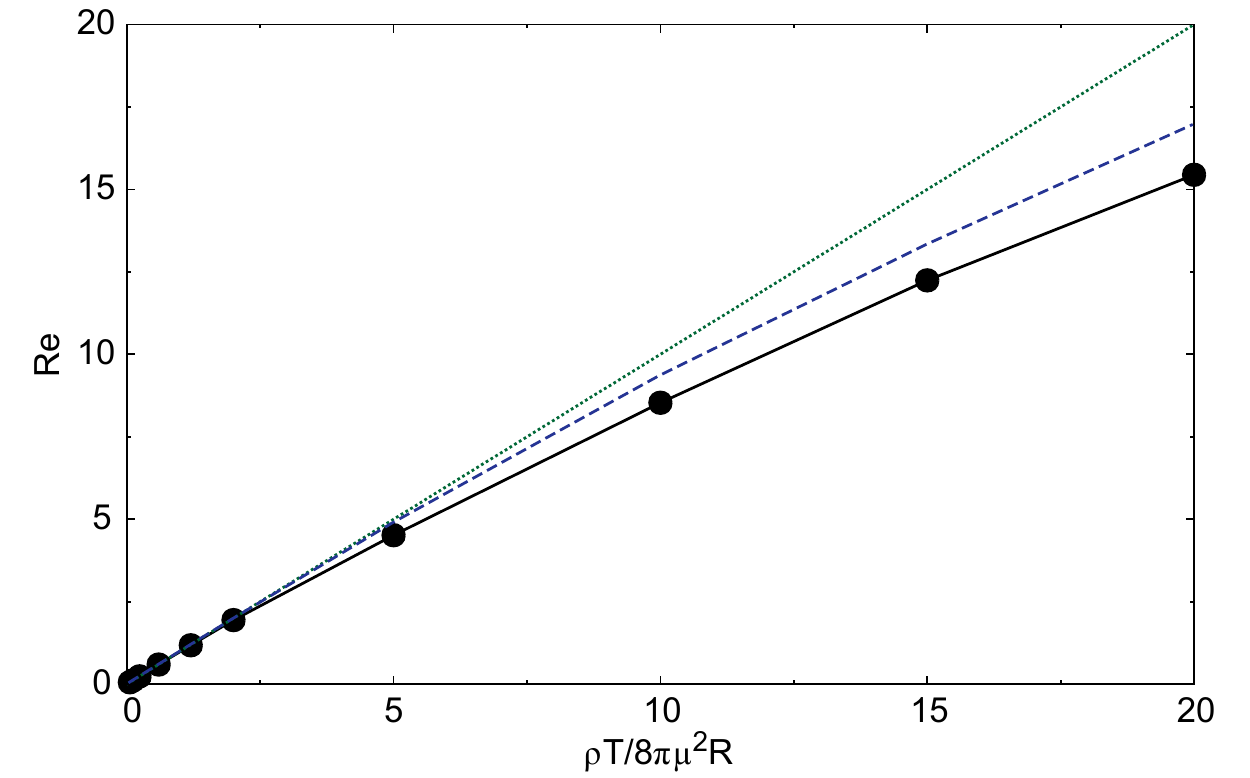}
\caption{\footnotesize {Validation of numerical method.} The measured Reynolds number from the simulations (black symbols) as a function of the theoretical predictions calculated using Stokes' limit $\omega_0 = T/8\pi\mu R^3$. At values $\mathrm{Re} \gtrsim 3$ the measured values start to deviate from the Stokes limit (green dotted line), following a prediction $T/8\pi \mu \omega R^3=1+\mathrm{Re}^2/1200-15647\mathrm{Re}^4/20744640000$ (blue dash line) as expected~\cite{bickley1938lxv,collins1955steady,liu2010wall}. Throughout this work the Reynolds number based on the Stokes limit $\mathrm{Re}=\rho \omega_0 R^2/\mu$ is used for clarity. The simulation box is $27R \times 27R \times 27R$ in this validation.}
\label{validation}
\end{figure}

In order to characterise the formation dynamics of the vortices, simulations with a periodic box of $8R\times 54R\times 54R$ were implemented (Fig.~\ref{phase}). Here, the formation of only a single particle vortex is observed (Fig.~\ref{phase}a).
In the steady state, the vortices span the periodic boundary along the vorticity direction, and the particles translate around the centre of the cluster with a tangential velocity $V_{\theta}(r)$ (Fig.~\ref{phase}a). The spinners entrain the  fluid leading to the formation of a liquid vortex with the same handedness as the particle one (blue arrows in Fig.~\ref{tornado}c). Inside the vortex, a solid body rotation $V_{\theta}(r)\sim r$ is observed for both the particles and the surrounding fluid ({\color{black} symbols and solid lines in Fig.~\ref{phase}b, respectively}). At the edge of the vortex the particle velocities drop slightly, while the fluid velocities start to decay (Fig.~\ref{phase}b). The particles exhibit random motion along the vorticity direction ($X$) showing diffusion-like dynamics (Fig.~\ref{phase}d and e), while a spiral motion is observed in the $YZ$ plane (Fig.~\ref{phase}c and Movie 2 in~\cite{SM}).

The normalised rotational frequency (the slope of $V_{\theta}(r)/\omega_0R$ in Fig.~\ref{phase}b) stays constant when the global volume fraction $\phi$ is increased (black and green symbols in Fig.~\ref{phase}b), but decreases when $\mathrm{Re}$ is increased (blue symbols in Fig.~\ref{phase}b). This implies that the energy conversion between particle rotation and translational motion is more efficient at lower Reynolds numbers. This is also manifested by the local volume fraction of the cluster (Fig.~\ref{phase}a): increasing the global volume fraction $\phi$ leads to a larger cluster at a constant $\mathrm{Re}$ (top three panels in Fig.~\ref{phase}a), while increasing $\mathrm{Re}$ with a constant $\phi$ results to less dense clusters (bottom two panels in Fig.~\ref{phase}a; see also Fig.~\ref{rho} {\color{black} for detailed density maps}).

\begin{figure}
\centering
\includegraphics[width=1\columnwidth]{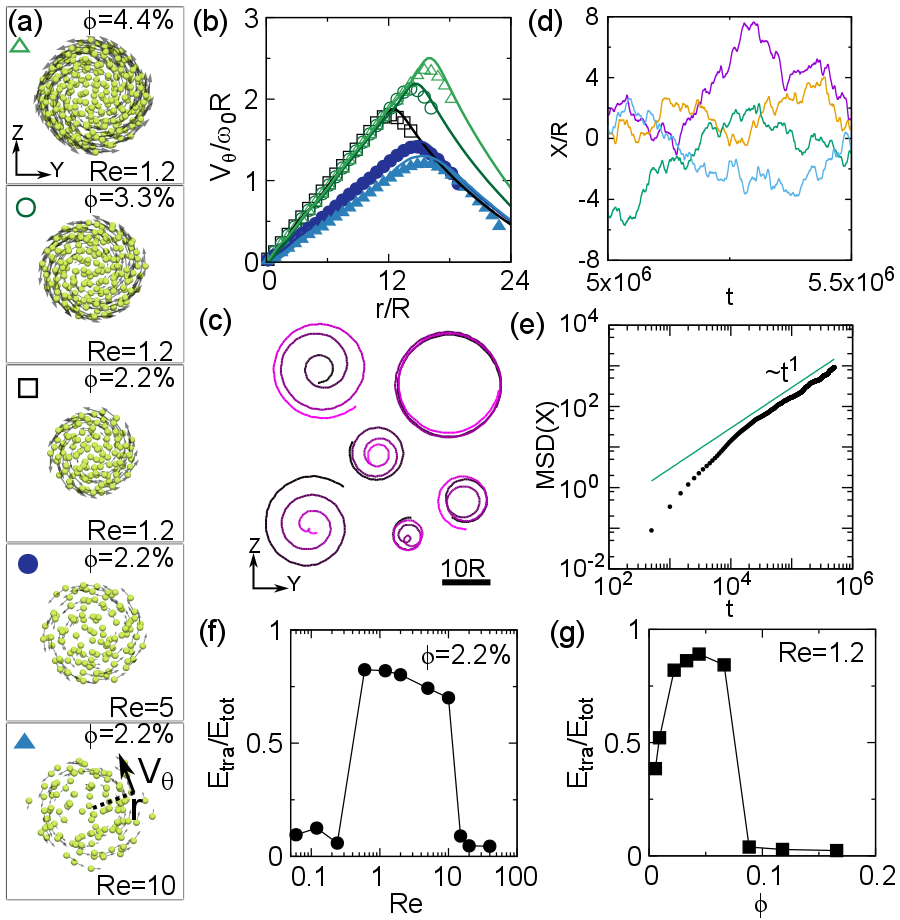}
\caption{\footnotesize {Phase diagram for the vortex formation in a thin periodic box ($8R\times 54R\times 54R$).} (a) Examples of the observed rotating clusters (The gray arrows show the velocity of the particles. The black arrow defines a tangential velocity $V_{\theta}(r)$ at a distance $r$ from the vortex centre.). (b) The tangential velocity $V_{\theta}$ as a function of the distance $r$ for different Reynolds numbers and volume fractions corresponding to the states in (a) ({\color{black} symbols and lines, for particles and fluid, respectively}). (c,d) Examples of the steady state particle trajectories in the vortex state ($\phi\approx 2.2$\% and $\mathrm{Re}\approx 1.2$) (c) in the $YZ$ plane perpendicular to the vorticity axis $X$ and (d) along the vorticity axis. (e) The mean square displacement (MSD) along the vorticity axis.
(f-g) The ratio between the translational and the total energies $E_{\mathrm{tra}}/E_{\mathrm{tot}}$ of the particles (f) as a function of the Reynolds number for $\phi\approx 2.2$\% and (g) as a function of the volume fraction $\phi$ for $\mathrm{Re}\approx 1.2$.
\label{phase}}
\end{figure}

\begin{figure}[tbhp!]
\centering
\includegraphics[width=1\columnwidth]{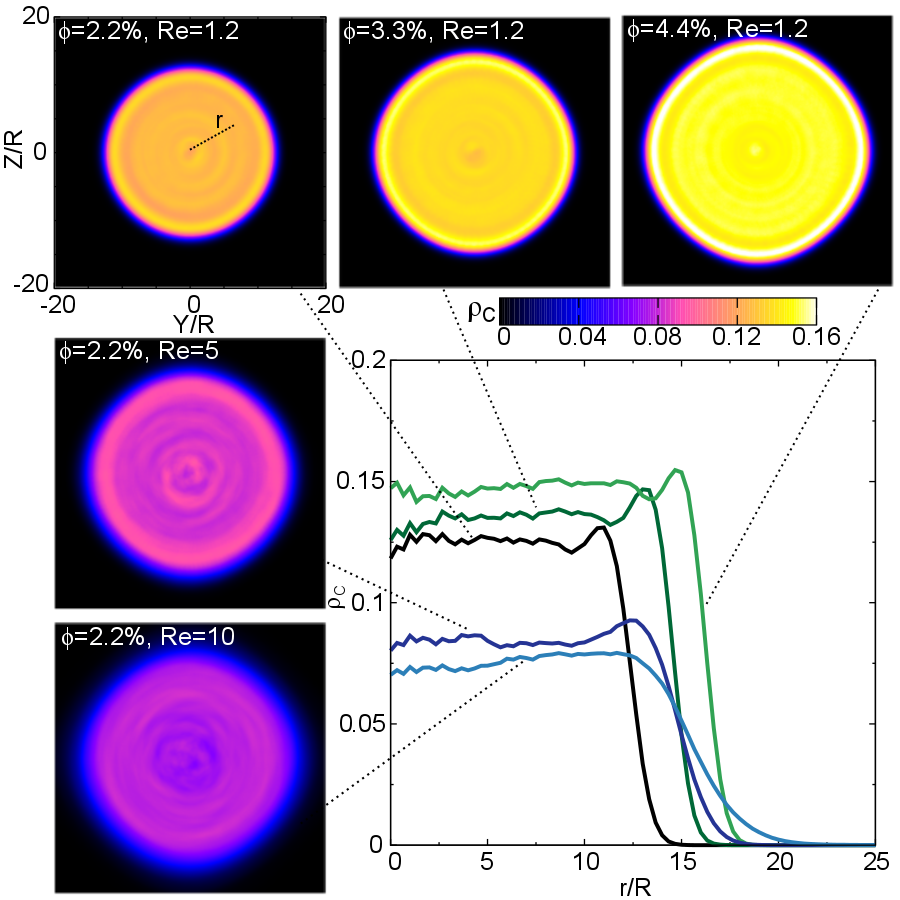}
\caption{\footnotesize The local particle volume fraction maps and the averaged local volume fraction $\rho_c(r)$  as a function of the distance $r$ from the centre of the vortex, in the plane perpendicular to the particle vorticity direction ($X$ axis), for various Reynolds numbers $\mathrm{Re}$ and global particle volume fractions $\phi$. For a steady state in the phase separation regime, we observe a rotating cluster with a local particle volume fraction $\rho_C\sim 0.1$ surrounded by a pure fluid depleted of the particles. Both the radial and azimuthal particle density are isotropic and reasonably constant inside the vortex with a small peak at the periphery.}
\label{rho}
\end{figure}

The drive $T$ gives a rotational energy input into the system, which is damped by a hydrodynamic friction. A total energy of the particles is defined as the sum of the rotational and translational energies: $E_{\mathrm{tot}}=E_{\mathrm{rot}} + E_{\mathrm{tra}}$. {\color{black} These are given by $E_{\mathrm{tra}} = \sum_{i=1}^{i=N}\frac{1}{2} mV_i^2$ and $E_{\mathrm{rot}} = \sum_{i=1}^{i=N} \frac{1}{2}I\omega_i^2 = \sum_{i=1}^{i=N} \frac{1}{5}m(\omega_i R)^2$. The $m$ and $I$ are the mass and moment of inertia of a spinner. The instantaneous velocity $V_i$ and the rotational frequency $\omega_i$ for a particle $i$ are measured from the simulations.}
When the vortex formation is observed, the total energy is dominated by the  translational motion of the particles. In a steady state a typical value  $E_{\mathrm{tra}}/E_{\mathrm{tot}}\sim 0.8$ is observed ({\color{black}Fig.~\ref{phase}f}). In the absence of the vortices,  the total energy $E_{\mathrm{tot}}$ is dominated by the rotation of the spinners ({\color{black}Fig.~\ref{phase}f}).

Using the ratio $E_{\mathrm{tra}}/E_{\mathrm{tot}}$ as an order parameter,
 we estimate the range of the  vortex formation in the thin samples, when either $\mathrm{Re}$ is varied for {\color{black}a} constant volume fraction $\phi$ or as a function of the $\phi$ for a constant $\mathrm{Re}$. When the volume fraction is fixed at $\phi\approx 2.2$\%, the vortex formation is observed for $0.3 \lesssim \mathrm{Re} \lesssim 10$ (Fig.~\ref{phase}f). Conversely, for a constant Reynolds number $\mathrm{Re}\approx 1.2$, a vortex is formed for  $\phi \lesssim 8$\% (Fig.~\ref{phase}g). For very low volume fractions $\phi\lesssim 2$\%, the translational energy rests finite, typically $E_{\mathrm{trans}}/E_{\mathrm{rot}}\sim 0.5$ (Fig.~\ref{phase}g), and the formation of small clusters is observed.

\subsection{{\color{black} Inertial effects}}
The spinning particles of radius $R$ are coupled to the surrounding fluid velocity field $\mathbf{u}$ by a no-slip boundary condition at the particle surface.
When the flow is slow ($\mathrm{Re}\approx 0$)
an isolated particle creates a rotating flow field with only an azimuthal component~\cite{climent2007dynamic,fily2012cooperative}
\begin{equation}
\mathbf{u}_{\theta}(r) = \left[\frac{\omega R^3}{r^2}\sin\psi +O(\mathrm{Re}^2)\right]\hat{\mathbf{e}}_{\theta}
\label{eq:az}
\end{equation}
where $\psi$ and $\theta$ are the polar and azimuthal angles in spherical polar coordinates. This leads to a constant rotational motion of a particle pair around a central point~\cite{climent2007dynamic,fily2012cooperative}. As it does not include attraction nor repulsion, it is not expected to lead clustering. Indeed, no clustering is reported in simulations of spinners at low $\mathrm{Re}$ and $\phi$~\cite{fily2012cooperative,yeo2015collective} in agreement with our simulations (Fig.~\ref{phase}f).

Expanding to small but finite Reynolds numbers gives rise to a secondary flow {\color{black} due to inertia}~\cite{bickley1938lxv,climent2007dynamic,liu2010wall}
\begin{eqnarray}
\mathbf{u}_r(r) &=& \left[-\frac{\omega R^3}{8r^2}\left(3\cos^2\psi -1\right)\left(1-\frac{R}{r}\right)^2\mathrm{Re}\right]\hat{\mathbf{e}}_r\label{eq:r}\\
\mathbf{u}_\psi(r) &=& \left[\frac{\omega R^4}{4r^3}\left(1-\frac{R}{r}\right)\sin\psi\cos\psi~\mathrm{Re}\right]\hat{\mathbf{e}}_\psi\label{eq:t}.
\end{eqnarray}
The radial component (eq.~\ref{eq:r}) creates an advection towards the particle in polar regions $|\psi | \lesssim 55^{\circ}$, while at the equator the fluid is advected away from the  particle (Fig.~\ref{fig:pair}a).
{\color{black} The effects  of the secondary flow has been attributed to the repulsion between two spheres spinning side-by-side~\cite{climent2007dynamic,aragones2016elasticity,steimel2016emergent} as well as to an attraction of a single spinner towards a flat wall along the vorticity direction~\cite{liu2010wall}.}

\begin{figure}
\centering
\includegraphics[width=1\columnwidth]{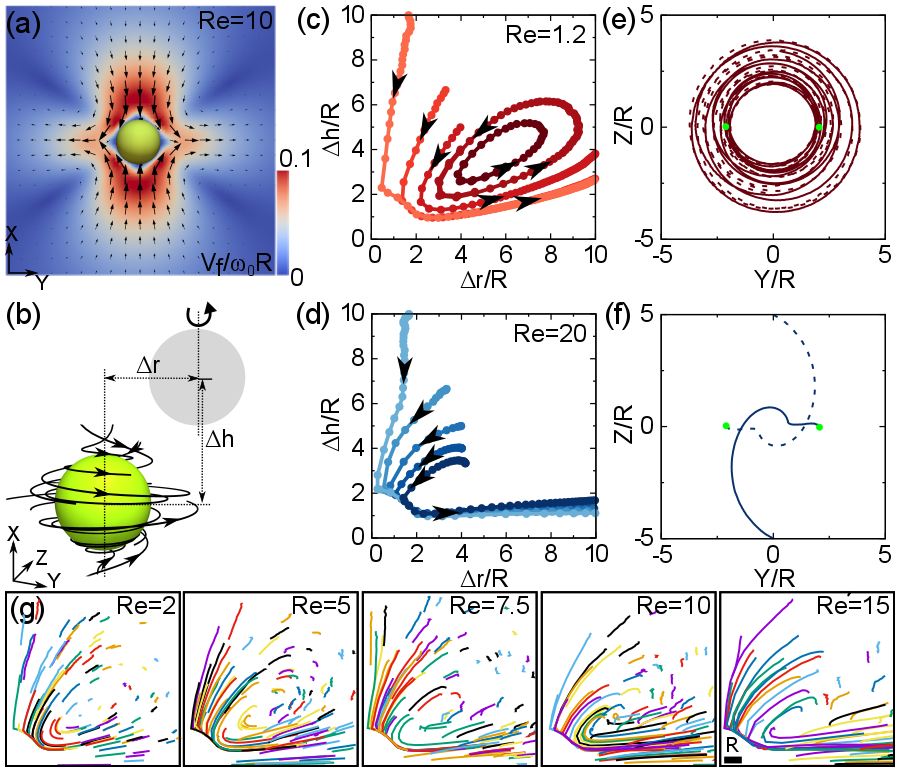}
\caption{\footnotesize {Hydrodynamic interactions of a spinners pair.} (a) The observed flow field of an isolated spinner at $\mathrm{Re}\approx 10$  at the particle mid-plane ($XY$) (the particle rotates around $X$ axis). (b) Schematic defining $\Delta r$ and $\Delta h$ between the two spinners. Observed trajectories of a particle pair in $\Delta r$-$\Delta h$ space: for (c) $\mathrm{Re}\approx 1.2$ and (d) $\mathrm{Re}\approx 20$ (simulation were carried out in a cubic periodic box of $40R\times 40R\times 40R$). Examples of the trajectories in $YZ$ plane perpendicular to the vorticity axis ($X$), for (e) $\mathrm{Re}\approx 1.2$ and (f) $\mathrm{Re}\approx 20$ for both of the particles in the pair (solid and dashed lines, respectively. The green dots mark the initial positions). {\color{black} (g) Observed trajectories in the   $\Delta r$-$\Delta h$ space, from short simulations ($t=20000 \Delta t$) for approximately 100 different initial conditions for different Reynolds numbers (the colours corresponds to different initial conditions).}\label{fig:pair}}
\end{figure}

{\color{black} The stabilisation of the clusters, requires the formation of structures via hydrodynamic interactions at $\mathrm{Re}\sim 1$.}
As a smallest possible building block,  we considered a spinner pair (Fig.~\ref{fig:pair}c-f). We measured the vertical $\Delta h$ and radial $\Delta r$ separations between the two particles (Fig.~\ref{fig:pair}b).
{\color{black} The trajectories show a general trend akin to the single particle flow fields: an attraction along the vorticity direction and repulsion at the equatorial region (Fig.~\ref{fig:pair}c and d). However, for a $\mathrm{Re}\approx 1.2$  the appearance of  a limit cycle in the $\Delta h$-$\Delta r$ space  is observed (Fig.~\ref{fig:pair}c; Movie 3 in~\cite{SM})). The inertial forces stabilise the particles into circular orbits around each other in the plane perpendicular to the vorticity (Fig.~\ref{fig:pair}e) while the $\Delta h$ and $\Delta r$ undergo periodic oscillations.
When the Reynolds number is increased, {\color{black} the inertial repulsion between the spinners increases and the limit cycle becomes less pronounced and eventually disappears: no hydrodynamic bound state was observed for $\mathrm{Re}>10$ (Fig.~\ref{fig:pair}d, f and g).} {\color{black}


\begin{figure*}[tbhp!]
\centering
\includegraphics[width=0.9\textwidth]{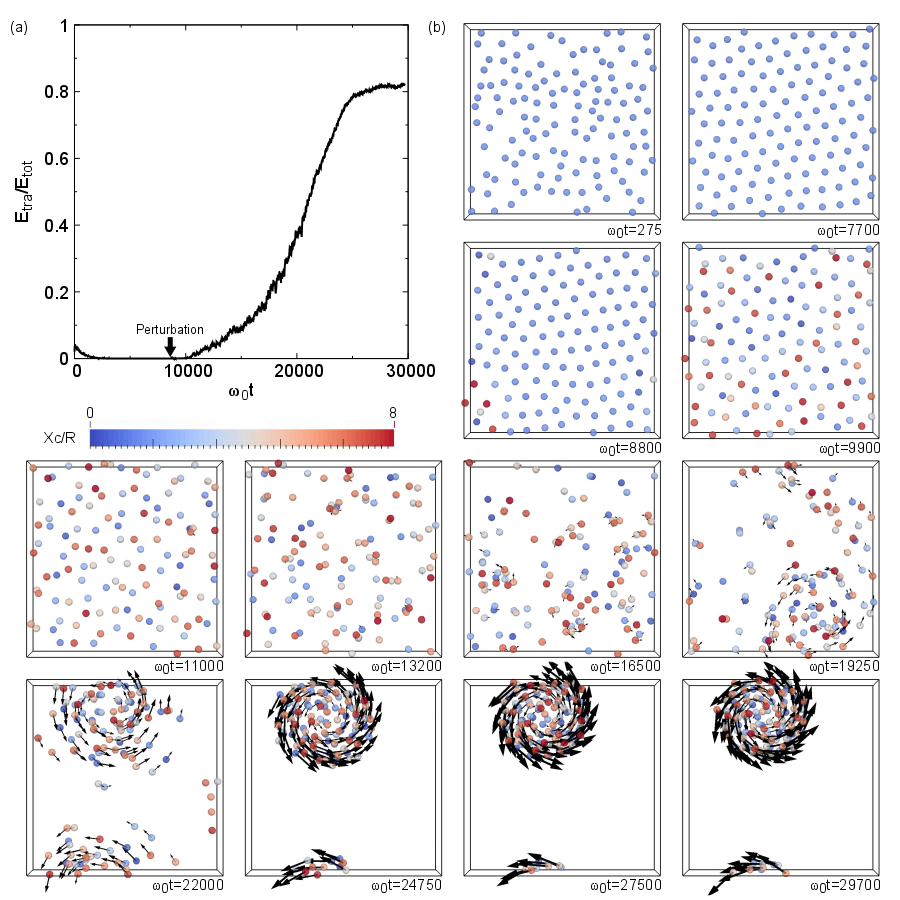}
\caption{\footnotesize (a) The time evolution of the ratio between the translational and the rotational energies $E_{\mathrm{tra}}/E_{\mathrm{rot}}$ and (b) snapshots of the configuration at different times. The simulations were initialised with the particles in a monolayer at $t=0$. In this case, the particles do not leave the monolayer and only experience in-plane hydrodynamic repulsion, leading to the formation of a 2-dimensional hexagonal structure at $\omega_0t=7700$. A small perturbation along the $X$ for 2 particles was introduced at $t\omega_0\approx 8000$. This leads to an onset of vertical particle motion, where the spinners start to explore the full 3-dimensional space (snapshots at $\omega_0t=8800$ and $\omega_0t=9900$).
The particle 3-dimensional hydrodynamic interactions lead to initially the formation of small and short lived clusters (snapshots at $\omega_0 t=11000$ to $\omega_0 t=16500$). Eventually,  the formation of a stable rotating cluster is observed at $\omega_0 t=24750$. (The simulations were carried with $\mathrm{Re}\approx 1.2$ and $\phi\approx 2.2$\% using a periodic simulation box $8R\times54R\times54R$).}
\label{xmove}
\end{figure*}

In the thin periodic simulation boxes, the formation of columnar vortices along the vorticity direction was observed (Fig.~\ref{tornado}) and
the particles have 3D trajectories inside the cluster (Fig.~\ref{phase}c and d). These suggest complex 3D flow effects and the importance of the interactions along the vorticity direction.
{\color{black} To highlight the importance of the 3rd dimension, we implemented a simulation where the particles were initialised randomly in a monolayer perpendicular to the vorticity axis (Fig.~\ref{xmove}). In this case no vortex condensation was observed: the randomly distributed particles stay as a monolayer and initially explore the 2D space due to the mutual advection from the azimuthal flow fields. The radial repulsion leads to the formation of a stable hexagonal spinner crystal with no translational motion $E_{\mathrm{tra}}\approx 0$ (see time $\omega_0t=7700$ in Fig.~\ref{xmove}a and b). At $\omega_0t\approx 8000$ a small perturbation of 2 particles along the vorticity axis ($X$) was introduced. This renders the monolayer unstable and the onset of vertical particle motion arising from the interactions along the vorticity direction is observed. The spinner trajectories become 3-dimensional and the formation of small hydrodynamically bound clusters is observed. Eventually, the hydrodynamic interactions lead to the formation of a stable spinner vortex (see $\omega_0t > 8000$ in Fig.~\ref{xmove}).}


It should be noted that a stabilisation of rotating clusters has been predicted to occur through hydrodynamic attraction between spinning disks in strictly 2D~\cite{goto2015purely}. However, when the third dimension is opened, the interactions become repulsive~\cite{goto2015purely}. Our results indicate that spheres spinning in 3D are markedly different. {\color{black} The single particle flow field gives rise to a hydrodynamic attraction along the vorticity direction and to a repulsion at the equatorial plane. {\color{black} The pair dynamics show more complex inertial effects. A hydrodynamic bound state is obtained for $\mathrm{Re}~\sim 1$ in the rotating $(\Delta r,~\Delta h)$-frame  (Fig.~\ref{fig:pair}c {\color{black} and g}). The particles move around a central point with a velocity corresponding to a translational Reynolds number $\mathrm{Re_T}\sim 0.1$}. Similar states are  also observed at early times in the bulk simulations (see {\it e.g.} Movie 1 in~\cite{SM}). {\color{black} The stabilisation of the large clusters likely relies on more complex inertial many body interactions in the presence of a large scale vortical motion. In the stable vortex state, both the rotational and translational particle Reynolds numbers are $\mathrm{Re}\sim 1$ (See Appendix A). The spinners do not have clear pair dynamics, but instead the particles have helical trajectories on the plane perpendicular to the vorticity axis, while diffusive behaviour is observed along the vorticity direction (Fig.~\ref{phase}c, d and e).}

\begin{figure}
\centering
\includegraphics[width=1\columnwidth]{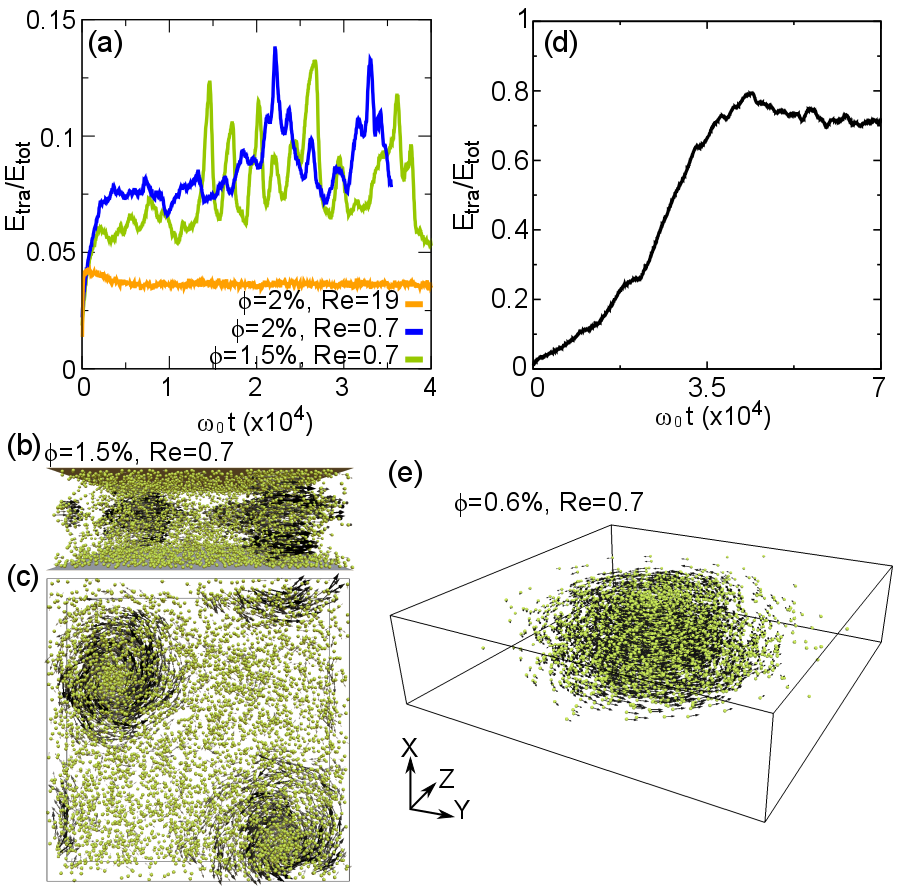}
\caption{\footnotesize {\color{black} (a-c) The vortex formation between two parallel walls.} (a) The time evolution of the ratio $E_{\mathrm{tra}}/E_{\mathrm{tot}}$ between the translational and total energies for the case of vortex formation ($\mathrm{Re}\approx 0.7$ and $\phi \approx 2\%$ (blue line) and $\mathrm{Re}\approx 0.7$ and $\phi \approx 1.5\%$ (green line)) and for isotropic case ($\mathrm{Re}\approx 19$ and $\phi\approx 2$\% (orange line)).
(b, c) A snapshot showing the cluster for $\mathrm{Re}\approx 0.7$ with $\phi \approx 1.5$\%: (b) side view and (c) top view. {\color{black} (d,e) An example of the observed condensation in a periodic simulation box: (d) The time evolution of the ratio $E_{\mathrm{tra}}/E_{\mathrm{tot}}$ and (e) a snapshot of the cluster.} (The  simulation box was $57R\times 172R\times 172R$ in a-c and $57R\times 228R\times 228R$ in d and e.)\label{wall}}
\end{figure}

\subsection{{\color{black}Role of boundaries:}}
To highlight the robustness of our predictions and exclude periodic effects along the (attractive) vorticity direction as a driving force behind the condensation, {\color{black} we carried out simulations using a large rectangular box 
with two flat walls in the $YZ$ plane perpendicular to the spinning direction  $X$, as well as with periodic boundary conditions  (Fig.~\ref{wall}).}
Similarly to the thin samples, a spontaneous condensation of particles rich and particle poor regions is observed {\color{black} for $\mathrm{Re}~\sim 1$ and $\phi \sim 1$\% (Fig.~\ref{wall}).}

When the spinners are confined (Fig.~\ref{wall}a),  a re-entrant dynamics where a large scale vortical motion appears periodically is observed (Fig.~\ref{wall}a, b and c; see also Movie 4 in~\cite{SM}).  In the case of high Reynolds number, the system rests isotropic (see orange line in Fig.~\ref{wall}a for $\mathrm{Re}\approx 19$ and $\phi\approx 2\%$ sample).
Spinning spheres have shown to be attracted to the no-slip surfaces perpendicular to the spinning direction~\cite{liu2010wall}. This can have an effect on the stability and the formation dynamics of the vortices. The ratio $E_{tra}/E_{tot}$, calculated for the particles {\color{black} far away from the confining surface} (wall-particle distance $>8R$) (Fig.~\ref{wall}a) is lower than in the case of periodic box (Fig.~\ref{wall}d) and  shows strong oscillations corresponding to the spontaneous appearance and disappearence of large scale vortices (blue and green lines in the Fig.~\ref{wall}a).
These suggest that our predictions should be readily observable by experiments of a density matched colloidal or granular spinner suspensions confined in a Hele-Shaw type geometries.

\section{Conclusions}
Using large scale numerical simulations we have studied a simple spinner system consisting of spherical particles spinning in a Newtonian fluid at weakly inertial regime in 3 dimensions. Our results demonstrate an unexpected  instability where the rotational dynamics of the colloidal spinners leads to a large spatial density and velocity variations of the initially uniform suspension. At low but finite Reynolds numbers and volume fractions we observe a spontaneous phase separation between particle rich and particle poor areas.
{\color{black} The formation of colloidal vortices rotating the same direction as the spinners is observed. The clustering is due to complex 3D flow fields and it requires a finite amount of fluid inertia. At early times, the simulations demonstrate the formation of small hydrodynamically bound clusters (early times at Movie 1 in~\cite{SM}; see also Fig.~\ref{fig:pair}c and Movie 3 for a bound state of an isolated spinner pair). More complex many body interactions then lead to the onset of collective motion and eventually to the formation of particle vortices.}

The predictions should be readily observable in any density matched spinner suspension, provided that hydrodynamic interactions are dominant. The simulations of the confined samples (Fig.~\ref{wall}) suggest that interesting experimental possibilities are provided by reasonably large PMMA particles in transformer oil rotating due to Quincke effect~\cite{lemaire2008viscosity} or millimeter sized particles with an embedded magnet in a rotating magnetic field~\cite{godinez2012note}, confined in a microfluidic chambers.

\begin{acknowledgments}
ZS and JSL acknowledge IdEx (Initiative d'Excellence) Bordeaux for funding, Avakas and Curta cluster for computational time and Hamid Kellay for discussions and careful reading of the manuscript.
\end{acknowledgments}

\appendix
\section{Calculation of the translational Reynolds numbers in the vortex state}
In the vortex state, a solid body rotation is observed where the particles translate around the vortex centre with a tangential velocity $V_{\theta} \sim r$ where $r$ is the distance from the vortex centre, as shown in Fig.~2b in the main text. For a particle inside the vortex, we can define a translational Reynolds number as $\mathrm{Re_T} = V_\theta l/\eta$, where $l$ is a characteristic length scale and $\eta$ is the kinematic viscosity of the fluid. Using a typical observed value $V_\theta \approx 1\omega_0R$ from the Fig. 2b in the main text, we arrive to a translational Reynolds number as $\mathrm{Re_T} = \mathrm{Re_R}\frac{l}{R}$, where $\mathrm{Re_R}\equiv\mathrm{Re}$ is the rotational particle Reynolds number used in the main text. Now we get a single particle $\mathrm{Re_T}\approx 1$ and $\mathrm{Re_T}\approx 12$ for a cluster of size $l\sim 12R$.

\medskip
\section{Supplementary movie captions}
Movie 1: The formation of 4  spinner vortices. The Reynolds number is $\mathrm{Re}\approx 1.2$, and the volume fraction is $\phi \approx 1.1$\%. The simulation uses  $N=240$ particles  in a rectangular box $8R \times 108R \times 108R$.

Movie 2: High time resolution movie of the internal dynamics of the particles in the cluster ($N = 120$ particles in a rectangular box $8R \times 54R \times 54R$, $\mathrm{Re}\approx 1.2$, and  $\phi \approx 2.2$\%). One of the particles is coloured purple to aid the visualisation. The blue points on the sphere is a mark of a fixed position on the particle surface to show the spinning of the particle.

Movie 3: The formation of hydrodynamically bound particle pair. ($\mathrm{Re}\approx 1.2$, in a cubic periodic box $40R\times 40R\times 40R$).

Movie 4: Vortex condensation between two parallel walls ($N = 6000$ particles in a rectangular box $57R \times 172R \times 172R$, $\mathrm{Re}\approx 0.7$ and $\phi \approx 1.5$\% with particle radius $R=2.1$).

\bibliography{ref}



\end{document}